\documentclass[journal]{IEEEtran}
\usepackage[utf8]{inputenc}
\usepackage{cite}
\usepackage{amsmath,amssymb,amsfonts}
\usepackage[ruled, vlined ]{algorithm2e}
\usepackage{graphicx}
\usepackage{textcomp}
\usepackage{xcolor}
\usepackage{tikz}
\usepackage{float}
\usepackage{url}
\ifCLASSINFOpdf
\else
\fi
\begin{document}
%
\title{Applying Intelligent Reflector Surfaces for \\Detecting {Violent Expiratory} Aerosol Cloud using Terahertz Signals }
%
%
%

\author{
Harun \v{S}iljak,~\IEEEmembership{Senior~Member,~IEEE,} 
Michael Taynnan Barros,~\IEEEmembership{Member,~IEEE,} 
Nathan D'Arcy,
Daniel Perez Martins,~\IEEEmembership{Member,~IEEE,}
Nicola Marchetti,~\IEEEmembership{Senior~Member,~IEEE,}
Sasitharan Balasubramaniam,~\IEEEmembership{Senior~Member,~IEEE}
\thanks{H\v{S}, ND and NM are with SFI CONNECT Research Centre, Trinity College Dublin, Ireland; DPM is with the SFI VistaMilk Research Centre, Walton Institute, Waterford Institute of Technology, Ireland; MTB is with the University of Essex, United Kingdom; SB is with the University of Nebraska, Lincoln, USA.
This publication has emanated from research conducted with the financial support of Science Foundation Ireland (SFI) and the Department of Agriculture, Food and Marine on behalf of the Government of Ireland under Grant Number [16/RC/3835] - VistaMilk research centre, and is co-funded under the European Regional Development Fund for the CONNECT research centre (13/RC/2077\_2).
Corresponding author (H\v{S}): harun.siljak@tcd.ie}
}

\maketitle

\begin{abstract}
The recent COVID-19 pandemic has driven researchers from different spectrum to develop novel solutions that can improve detection and understanding of SARS-CoV-2 virus. In this article we propose the use of Intelligent Reflector Surface (IRS) and terahertz communication systems to detect {violent expiratory} aerosol cloud that are secreted from people. Our proposed approach makes use of future IRS infrastructure to extend beyond communication functionality by adding environmental scanning for aerosol clouds. Simulations  have also been conducted to analyze the accuracy of aerosol cloud detection based on a signal scanning and path optimization algorithm. Utilizing IRS for detecting {violent expiratory} aerosol cloud can lead to new added value of telecommunication infrastructures for sensor monitoring data that can be used for public health.
\end{abstract}


%
\IEEEpeerreviewmaketitle

\section{Introduction}
%
%
%
%

\IEEEPARstart {I}{n} the fight against the COVID-19 pandemic, various disciplines have come together to develop solutions that will help in not only eradicating the virus, but also characterizing its propagation properties to curb the pandemic. The information communication technology (ICT) research community has also contributed to this effort; this includes developing new contact tracing tools and the use of artificial intelligence (AI) to assist in mass surveillance systems.  Having knowledge of whom an infected individual may have exposed to {virus}  allows health agencies to selectively quarantine people at risk and halt the further spread of viral infection. This paper joins this global effort by using 6G wireless network infrastructure to detect sources of  viral outbreaks that can help curb pandemic spreading.

{Violent expiratory events} such as  sneezing  generate vast amounts of aerosols and large droplets. As an example, recent studies have shown that the dominant way of SARS-CoV-2 are the aerosol transmissions \cite{Tang21}. While it would not be possible to sense the presence of the virus in the air directly, we make the case that detecting the major {violent expiratory events} is important in indoor spaces during an epidemic of a virus whose transmission mode is aerosol.

\begin{figure*}[ht!]
    \begin{center}
        \includegraphics[width=1.9\columnwidth]{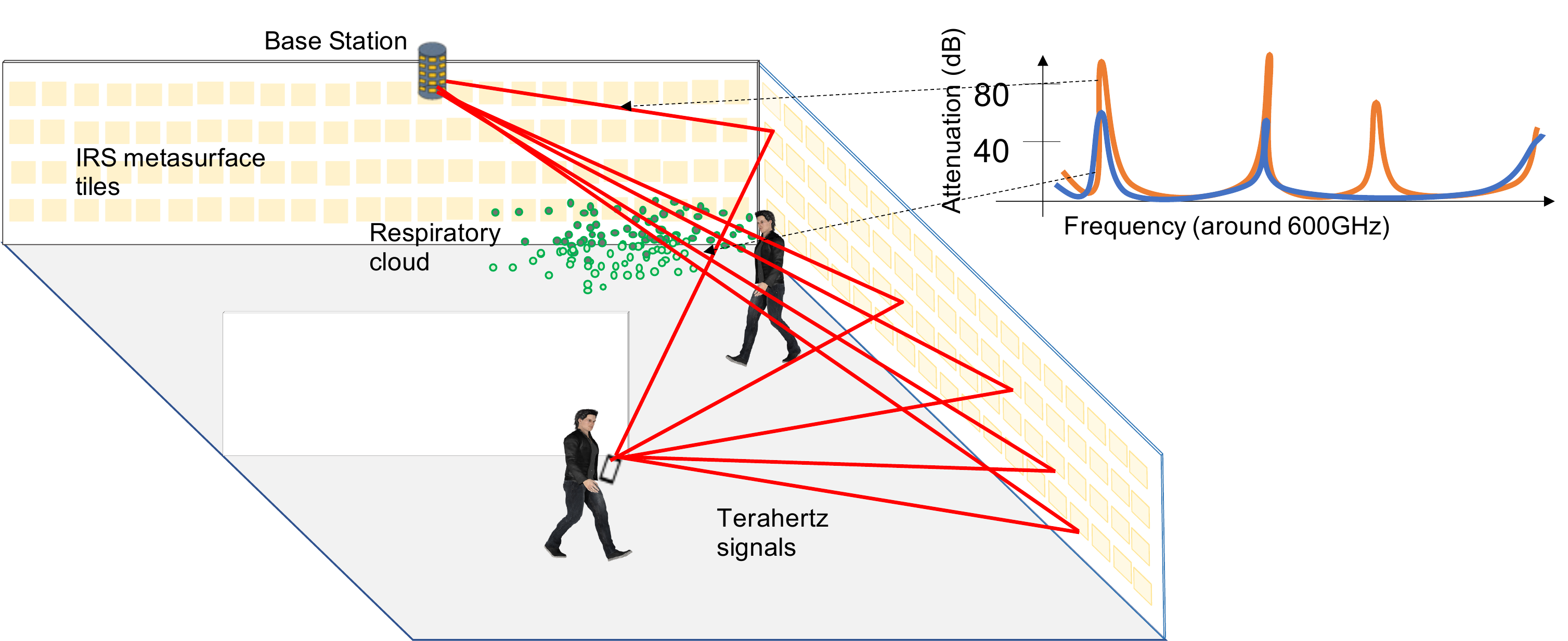}
    \end{center}
    \caption{The signals that are reflected from the IRS will  criss-cross specific regions of {violent expiratory} cloud, where its attenuation is used to determine the location and spreading patterns of the airborne droplets. As shown in this figure, this is based on the terahertz signals that have high molecular absorption factor that leads to signal attenuation.}
    \label{fig:Fig1}
\end{figure*}

We present a solution using terahertz wireless signals to detect {violent expiratory events} (e.g., sneezing) which, in case of an airborne virus infection significantly increase the viral particle concentration in the air. This information can be utilized by public health officials to understand the spreading pattern of an infection, provide guidance for ventilation, or improve contact tracing accuracy. Our strategy exploits the current 6G research  trends in utilizing \emph{Terahertz} frequency spectrum for sensing, and in particular this spectrum's unique molecular absorption properties. Given the sensitivity of  the terahertz  spectrum  that requires Line-of-Sight (LoS), the research community has proposed a new infrastructure called \emph{Intelligent Reflector Surfaces} (\emph{IRS}) that will bounce and intelligently re-direct signals from the walls, 
allowing a degree of control over the wireless propagation environment. These consist of metasurfaces integrated
and controlled with programmable electronics and allow for total control over the direction, polarization, amplitude, and phase of impinging waves \cite{liaskos2018new}. This means that an indoor environment can be covered by intersecting wireless thin beam rays (as shown in Figure  \ref{fig:Fig1}), providing an opportunity to track and localize respiratory vapor cloud by monitoring spatial signal attenuation that are reflected from IRS.

{Joint communication and sensing is cited as a major driver of THz communications in 6G. \cite{chaccour2022seven} Sensing capabilities can come from bandwidth exploitation: examples include radar-like sensing of rigid objects and localization of user equipment. The other source of sensing power is the molecular absorption we mention earlier: however, the known applications of THz sensing like rotational spectroscopy (e.g. for detecting ethanol in human breath) aim at scanning small, enclosed samples of gas. Moreover, oftentimes the reference to THz sensing capacity actually refers to THz network's ability to communicate with nanosensors, thanks to its short wavelengths, and not actually using the THz wave as a sensing vector. In our work, we envision a large scale molecular absorbtion-based sensing scheme coexisting with communication.}

In this paper we demonstrate how scanning of indoor areas for clouds generated by {violent expiratory events} using terahertz signals and an unused portion of tiles of the IRS can be seamlessly integrated with their main role for wireless communications. {While this particular work focused on IRS, the tracking of aerosol clouds can also be done via Reconfigurable Intelligent Surfaces, such as HMIMOS \cite{huang2020holographic}, in order to integrate transceivers emitting signals that penetrate through the aerosol cloud.} Through a simulation campaign we demonstrate the reliability of such a system, and we discuss the challenges, future directions, and effects this approach would have on future communication systems. The objective of this paper is to sense respiratory cloud, and not any specific virus. Rather, our aim is to determine a mechanism of detecting {violent expiratory events} using telecommunications infrastructure, as a new tool to help detect {sneezing} vapor that results in a new pandemic. The wavelength at THz frequencies allows the necessary sensitivity of the signal to detect aerosol presence; the future pervasive deployment of THz and IRS infrastructure allows widespread use of our solution without an additional installation cost, sharing unused infrastructure and spectrum whenever possible.




\section{Airborne Viral Propagation}

{COVID-19 has predominantly spread between people through a number of mechanisms, including airborne transmission \cite{Tang21}. Through this propagation method, viral particles emitted through coughing or sneezing form high-propulsion respiratory clouds that can disperse from few centimeters up to some meters, depending on the room conditions \cite{Li21}. Airborne propagation is highly impacted by the evaporation levels within the air. This is also dependent on the size of the droplets, where usually sizes are approximately 8 - 16 $\mu$m, but the highest viral propagation exists for droplets that are of the size of 32 - 40 $\mu$m. In the case where there is no evaporation, the propagation of the clouds can be for a few meters at high density where the airborne viral particles can sit in the air for up to 10 seconds. This can lead to airborne transmission for environments with low humidity conditions.}

{In addition to the physical modeling of airborne viral propagation, recent research studies have been conducted in understanding how virus propagate through the air using molecular communications theory \cite{Khalid19}, \cite{barros2021molecular}. For instance, Khalid et al. modeled the emission and reception of airborne molecules as a communications systems while Barros et al. conducted a thorough analysis of viral propagation due to breathing, coughing and sneezing molecular communication models and data \cite{Khalid19}, \cite{barros2021molecular}. These models utilize the physics of airborne viral propagation to represent the communications processes that occur between a person emitting a high-propulsion exhalation and another person receiving this airborne viral concentration. Please note that these models consider several parameters, such as the distance between transmitter and receiver, the emitted airborne viral concentration, free space propagation channel characteristics, and the mouth as a cone antenna that gives direction to the emitted signal. We build our simulation scenarios (see Figure \ref{fig:my_label}) based on such molecular communications models.}

\section{Terahertz Signals and Intelligent Surface Reflectors}

\subsection{Terahertz Signals Properties}

Terahertz electromagnetic waves suffer  from high free space path loss and molecular absorption losses, creating a requirement for near LoS communication \cite{polese2020toward,chen2021terahertz}. The most significant and severe absorption losses are due to water vapor \cite{jornet2011channel}. Instead of looking at the molecular absorption properties of terahertz waves as a negative, our approach exploits this signal attenuation as a tool for {violent expiratory} aerosol cloud detection.

The attenuation plot in Figure \ref{fig:Fig1} illustrates the effect of temperature and humidity on molecular absorption loss of electromagnetic waves at frequencies of our interest. 
The resonance pattern for water vapor in the terahertz spectrum produces windows where the absorption coefficients are small, and windows where the absorption coefficients are large, i.e., the channel is frequency selective. 
There is a stark difference in attenuation a wave traveling through a respiratory cloud suffers in comparison to a wave propagation through standard room conditions. 

By transmitting a signal through a region of space and measuring the received signal power at the receiver as illustrated in Figure \ref{fig:Fig1}, our aim is to deduce whether that signal has passed through an increased water vapor concentration. Using the beam-forming and reflection capabilities of the IRS, the multiple thin beam rays can be used to scan all areas of an indoor environment. {These rays look for a sudden increase in {violent expiratory} cloud vapor concentration as illustrated in Figure \ref{fig:Fig1}}. This respiratory vapor cloud can then be tracked as it evolves over time, and user devices can be informed of its location. 
Using the received power measurement to detect the presence of a {violent expiratory}  vapor cloud requires no alteration of the information-bearing signal and the communication protocol in place. The service would piggyback on normal wireless communications.

\subsection{Intelligent Reflector Surfaces}


IRS is a technology that exploits the different dielectric properties of materials to reflect electromagnetic waves to enhance the performance of indoor wireless communications systems by countering the higher path and molecular absorption losses of THz signals \cite{liaskos2018new,huang2020holographic,temmar2021analysis,huang2021multi}.
The walls in a typical office room can be covered by reflective materials that can intelligently control the properties of a THz signal (e.g., amplitude and phase) so all users have access to efficient and high data rate wireless communications links. 

A wide variety of reflective materials (i.e., metamaterial) have been utilized as the main component for the fabrication of IRS. For example, Bi et al. \cite{bi2019experimental} have proposed the use of flexible ceramic microspheres that can reflect up to $95\%$ of the incident electromagnetic wave, with a bandwidth of $0.15$ THz, and obtaining resonance from $0.3-1.4$ THz depending on their physical configuration. In a different approach, Temmar et al. \cite{temmar2021analysis} have selected graphene as the IRS fabricating metamaterial, achieving a high gain ($\geq 8.6$ dB) for $0.6-0.64$ THz frequency range and a channel capacity of $29.10$ bits/Hz for a $(2\times2)$ MIMO configuration. 

\begin{figure*}
\begin{center}
        \includegraphics[width=15cm]{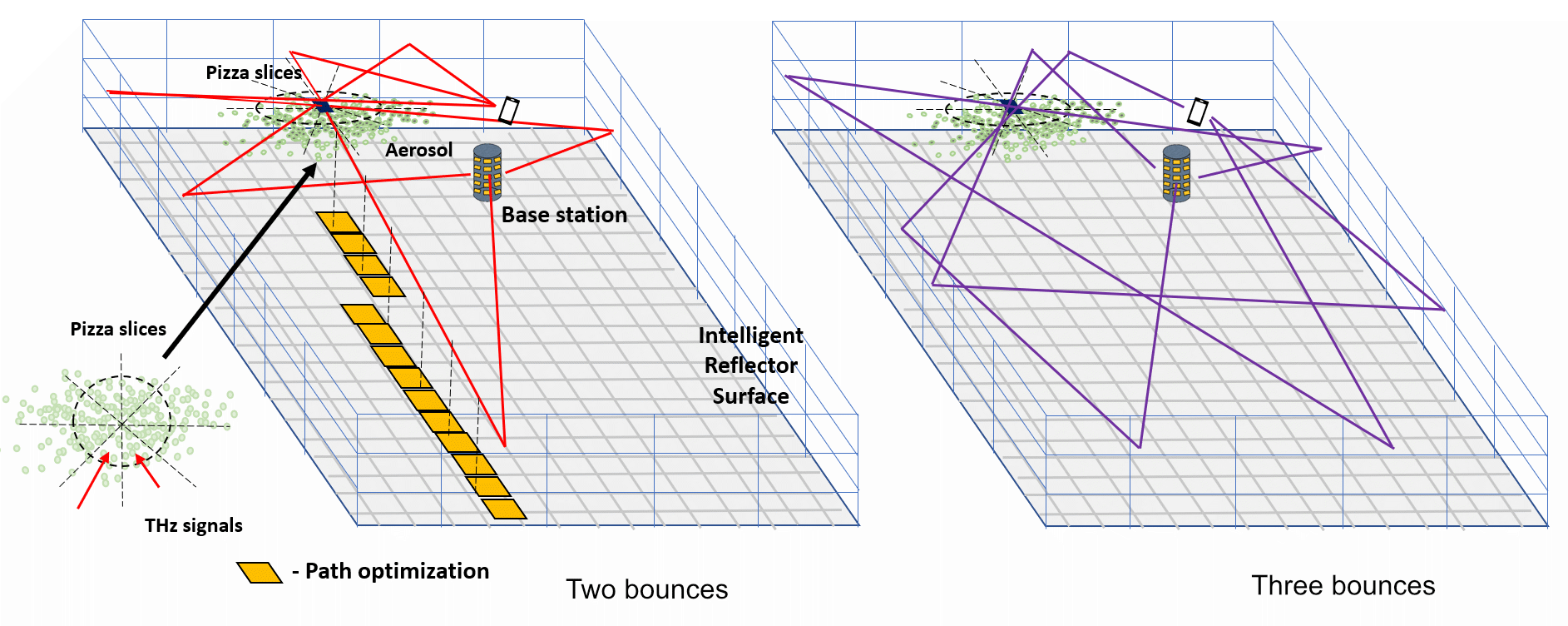}
    \end{center}
    \caption{Representation of the different propagation paths employed to scan the grid in the centre of the {violent expiratory} cloud (two bounces, i.e. ray consisting of three line segments, and three bounces, consisting of four line segments). Diverse paths produced by pizza slice approach are shown, alongside the intuition for path optimization.}
    \label{fig:my_label}
\end{figure*}

\section{Scanning for the {Violent Expiratory Aerosol} Cloud}

An important question for our approach is \emph{"How to efficiently check a room for the presence of {violent expiratory} cloud using a THz transmitter and a group of IRS on the room walls?"} {Our solution to this problem is shown in Figure \ref{fig:my_label} and it was created using Python.} We start by digitising the room layout and dividing it into a grid, creating a pixelated image of the room. 
The field-of-view of an IRS depends on the geometry of its construction and, therefore, it is not guaranteed to
be able to reflect an impinged signal with a full 180 degrees of freedom. This is mirrored within the simulation environment where a non-ideal field-of-view of 140 degrees is assumed. The idea of using multiple IRS in the indoor communication scenario is to avoid blockages by bouncing the signal between several IRS on its way between the base station and the user. By increasing the number of 
reflections in a path, signals can also take more unique and distinct trajectories around the room. Two different propagation path types will be considered which utilize differing amounts of
these reflections; two and three bounces as indicated in Figure \ref{fig:my_label}. 


\begin{figure}[tb]
    \begin{center}
        \includegraphics[width=8cm]{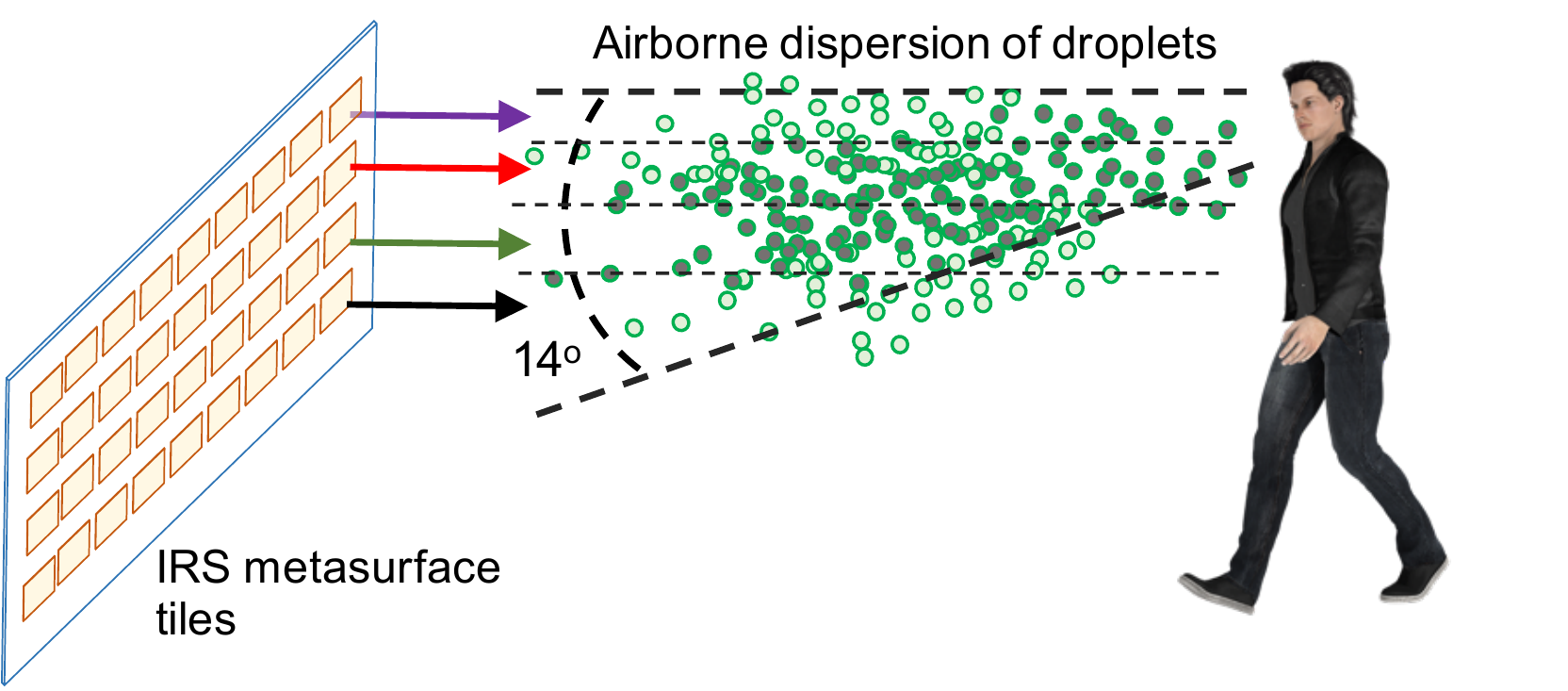}
    \end{center}
    \caption{The dispersion of droplets through the air can take between 1 - $14^{\circ}$ in environments with no evaporation (after an initial expulsion of $27.5^{\circ}$) \cite{Li21}.}
    \label{fig:Devices}
\end{figure}

When a grid (a pixel in our discretized layout) is selected to be scanned, a limit on the amount of different signal paths to be used has to be set as it is infeasible to use all IRS in the room to scan a single grid. The best paths available are also chosen to be used. It is preferable for the rays which intersect the grid to come
from significantly different trajectories, travelling through different sections of the room. These paths are chosen using the \emph{Smart path} selection technique by first considering the unit circle around the grid to be
scanned. Depending on the maximum paths chosen, the 180 degree angle around the
selected grid is segmented. For example, consider a scenario where a maximum of three paths are chosen. The
180 degree angle around the grid is split into three equal segments as if it has been cut into three
equal pizza slices, each slice being 60 degrees. Within each pizza slice, a ray path is chosen that
enters the grid from within that 60 degree segment. As there will be multiple IRS that have paths
within each slice, the IRS and path with the shortest propagation is chosen to minimise
path loss.
This results in the selected grid being scanned from IRS evenly distributed around its unit
circle, harnessing IRS from all corners of the room. Allowing a greater amount of maximum IRS
increases the amount of slices/segments, and hence increases the resolution of how the grid will
be intersected from multiple signal beams.

As seen in Figure \ref{fig:Devices}, each row of metasurface can emit horizontal signals that will sense a particular height of the droplets. The 2D propagation will basically scan the room in a 2D plane, but this scanning process will move down the IRS tile by tile vertically. This effectively means that we will end up with a 3D scanning process. For each signal path, two received powers will be
calculated: a control power, and an actual received power. These two powers will be compared to
each other and a binary decision made on whether a selected grid contains a respiratory cloud. First, a control received power is calculated where it is assumed that there is no {violent expiratory} 
cloud present in the room. 
For the same ray path, its received power is calculated using the actual simulation room’s
data. For each grid the ray passes through, an absorption coefficient for that grid is calculated using
baseline temperature and relative humidity data. All the grids on the ray’s path are tallied. Using this tally, the ray’s total propagation distance can be divided into segments proportional to the grid conditions it has passed through. If the actual received power for each ray used to scan the grid is lower than its complement control received power, then the selected
grid is classified as having a {violent expiratory} cloud present. 

The described approach allows us to (1) check a particular grid for the presence of {violent expiratory event} cloud, and (2) diversify the rays in space. 
The simulations we will present in the remainder of the paper are small enough to allow for exhaustive search, checking all grids. In larger spaces, practical considerations ask for smart, guided grid selection and path optimization process. Here, experience from uncertainty modeling, mapping, or image processing, with associated tools such as Markov random fields can be helpful. Knowledge about grid's neighbours affects the expectation of the respiratory cloud in a particular grid, and the grids with higher expectation can be prioritised. Alongside this probabilistic approach, we use grid trail elimination as well. {There, if a given ray's received power matches the expected power, grids along the ray are marked as currently unaffected by the violent expiratory event cloud.} This reduces the search space for grids that need scanning.


\section{Simulations}

{Our study is based on simulation that builds on established and validated models that have proven path loss is affected by molecular absorption of water vapour. Our aim is to build on these validated path loss models to further investigate a new application of detecting violent expiratory events.}

\subsection{Simulator Development}



We constructed a 2D indoor simulation environment scenario of room size 10 m by 10 m, split into 10 cm by 10 cm grids. Every grid element holds information on that specific area of the room including
its temperature, humidity, and water vapor concentration. 

A base station is added to the centre grid of the room, and acts as the transmitter for
all signals. A network user’s position can be specified as an input, and its location placed in the
respective grid to act as the signal receiver. The IRS are placed on the walls of the room via these grids. The IRS locations and total coverage on the walls can, therefore, be varied when initialising the grids and the simulation environment. {Violent expiratory event} clouds are modeled by adjusting the water molecule concentration, temperature, and humidity in these grids. Using these properties we simulate the {violent expiratory} cloud propagation (using the algorithm proposed by \cite{Gulec2021}) considering its physical properties, such as air (environment and the fraction that leave the person's mouth) and droplets densities, as well as the air viscosity. Through this method, we also determine the number of droplets in the cloud.

With this simulation environment and the propagation loss model, ray-tracing methods were used to model the straight line propagation paths of transmitted terahertz signals and record the received signal power, considering free space pathloss. 
The scanning algorithm was designed to allow any particular location, i.e. a grid, in the room to be scanned for the presence of a {violent expiratory event} cloud. When a grid is chosen, the ray propagation paths are formulated from the base station to the user device which pass through the grid. The idea is to be able to have a selection of multiple possible ray paths, from the base station, reflected from the IRS, which arrive at the end user. These ray paths must crisscross or intersect the selected grid to be scanned. A number of these paths can then be chosen and signals transmitted along them. 

The performance of the scanning algorithms in detecting a {violent expiratory} cloud was measured using binary classification metrics. This consists of \emph{precision}, which is the probability that a grid that is classified as having a {violent expiratory} cloud present, and \emph{accuracy}, which is the probability that a classification produced by the scanning algorithms is
correct.

The effect of varying the following parameters on the performance of the scanning algorithm was examined: (1) the percentage IRS coverage of the walls of the indoor room, (2) the type and amount of signal path used to scan a single grid, and (3) the effect of user and {violent expiratory} cloud location. Operating frequency of 0.6 Thz was used based on a recent work in \cite{Deal17}, which demonstrated terahertz signal of 666 GHz achieving a peak data-rate of 9.5 Gbps at a distance of 590 meters. 
\subsection{Results}
\subsubsection{IRS Coverage}


Experiments were  conducted with a varying percentage of wall area covered by the IRS. 
Scanning simulations were executed for percentage IRS cover from 10\% to 100\%. 
For every grid, a maximum of 10 ray paths were chosen to scan that grid, with a maximum of 5 of both propagation path types (two bounces versus three bounces).

\begin{figure}
    \centering
    \includegraphics[width=\linewidth]{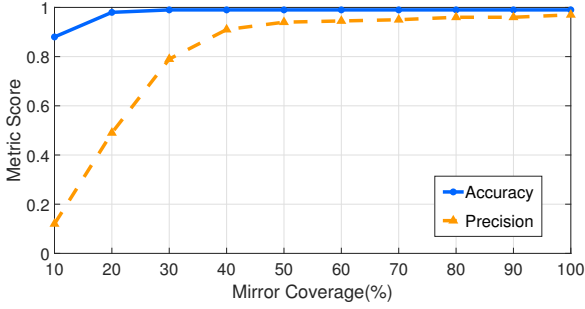}
    \caption{Metric scores versus IRS coverage on the walls of the indoor room.}
    \label{fig:Mirrors}
\end{figure}
The results presented in Figure \ref{fig:Mirrors} indicate that an exhaustive deployment of IRS in an indoor
room is not needed for the scanning system to perform well. The scanning system
performs nearly just as well with 50\% coverage as with 100\%. This reduces a hardware barrier to
implementing such a scanning system. Below 30\% coverage, the performance of the scanning algorithm exponentially decreases. 

\subsubsection{Signal Paths}

\begin{figure}[H]
    \centering
    \includegraphics[width=\linewidth]{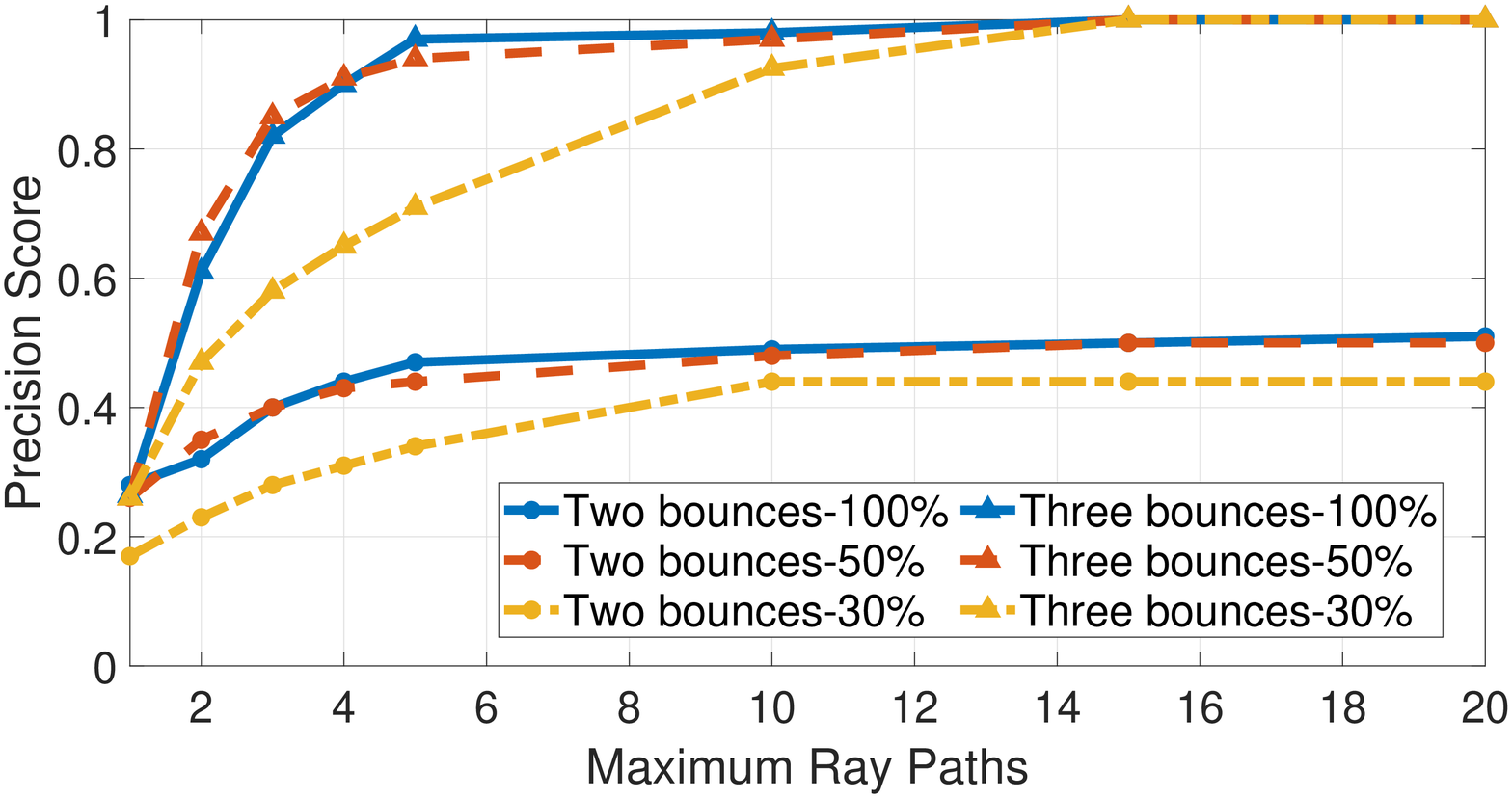}
    \caption{Precision score versus maximum number of two-bounce or three-bounce ray paths per scanned grid. Results are presented for an IRS coverage of 100\%, 50\%, and 30\%. }
    \label{fig:precision3}
\end{figure}

Results in Figure \ref{fig:precision3} reveal the optimal number of rays needed to scan a
single grid. We also compare the performance of each path
type, with varying allowable numbers of ray paths per scanned grid. Averages from experimental iterations over percentage IRS coverage of 100\%, 50\%, and 30\% were calculated, and the entire room was scanned.

As expected, at lower IRS coverage and lower number of maximum rays per grid scan, the
performance of the scanning algorithm decreases. Interestingly, at a IRS coverage of 30\% the
performance drops significantly below 10 maximum rays per grid scan. However, at an IRS coverage
of 50\% and 100\% the performance remains relatively stable until the maximum rays per grid
drops below 5. This would indicate that having a higher percentage IRS coverage in an indoor
room would allow for less ray paths to be used. A larger pool of
IRS in the room allows for a better selection of ray paths, and hence better performance with less
rays. However, having greater IRS coverage in a room presents a problem with
the physical and monetary barrier to installing them. Not every material can be installed with an
IRS, e.g. glass windows. On the other hand, using more ray paths when scanning will require
more signals to be transmitted from the base station, leading to higher power consumption
costs. While three-bounce paths outperform two-bounce paths, it must
be highlighted than three-bounce paths have a longer propagation distance and will hence suffer higher
path losses. To utilize three-bounce paths, antenna gain must be adequate to combat these
larger losses.

\subsubsection{User and {Violent Expiratory} Cloud Location}

\begin{figure*}[h]
    \centering
    \includegraphics[width=15cm]{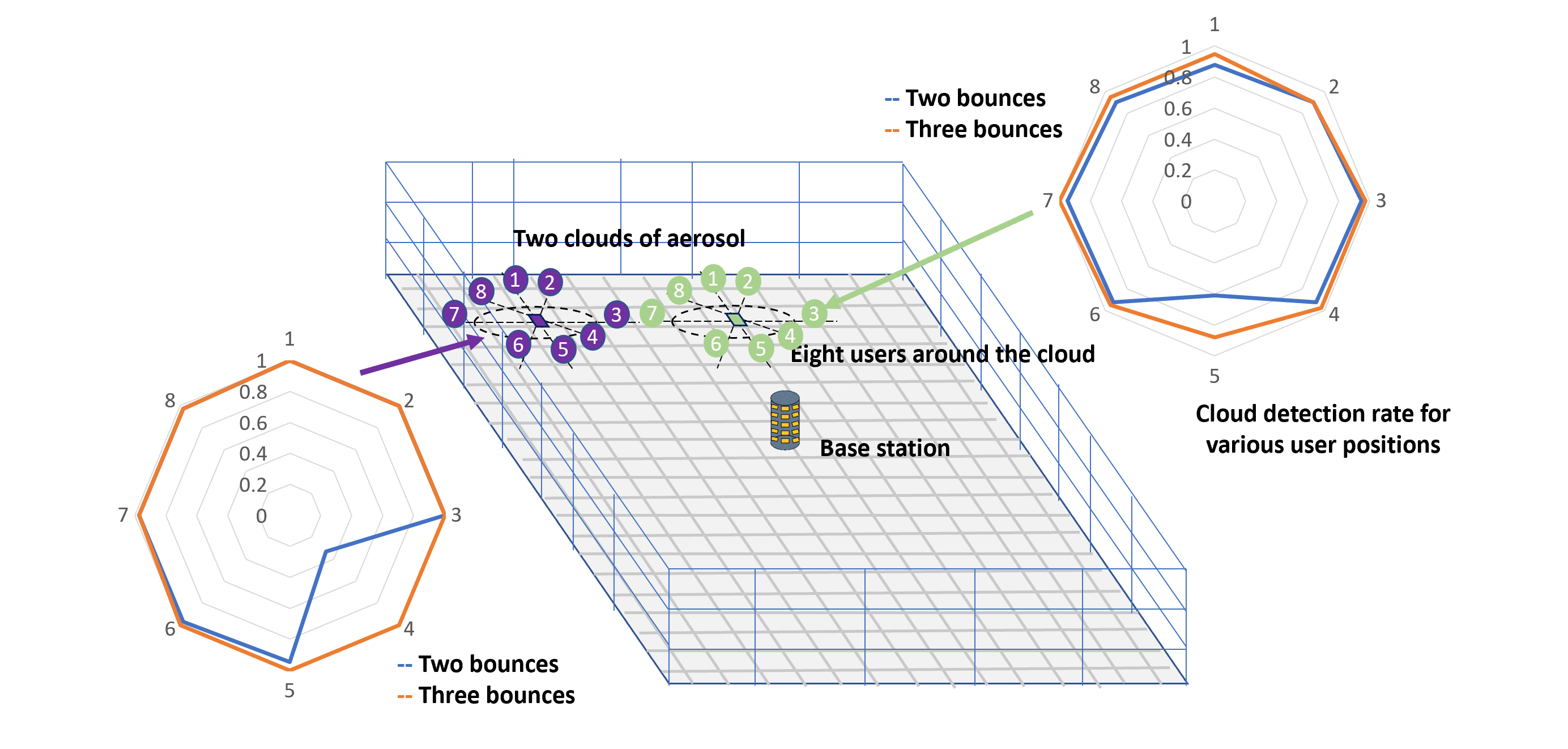}
    \caption{Two characteristic options for {violent expiratory} cloud position (diagonal and orthogonal), with 8 different positions of users around them. The spider charts correspond to the precision of detection w.r.t. user position around the cloud.}
    \label{fig:diag}
\end{figure*}

The effect of both the user and {violent expiratory} cloud location on the performance of the scanning is presented. Two characteristic cloud locations were considered and are displayed in Figure \ref{fig:diag}. The first cloud location lies to the diagonal of the base station. The second cloud location is above the base station, along its $y$ axis. A {violent expiratory}  cloud, in general case, will lie either to the diagonal of the base station, along its $x$ or $y$ axis, or somewhere in between these two cases.
At each of the two {violent expiratory} cloud locations, simulations were executed for eight different permutations of user location. The user locations denoted 1 through 8 in Figure \ref{fig:diag} lie at a distance of $1m$ around the respiratory cloud. From the spider charts the scanning algorithm can be seen to slightly under-perform the second cloud position but not significantly. This is due to the second cloud location being closer to the base station, effectively cutting off the base station's field-of-view with no {violent expiratory} cloud. The closer the {violent expiratory} cloud is to the base station, the greater the probability for a scanning ray to pass through the cloud, irrespective of the specific grid that is being scanned. 
This suggests that the position of the base station in the indoor scenario impacts significantly on the performance of the scanning algorithms.

There is one user location which results in a significant performance drop for both cloud locations when two-bounce paths are used by the scanning algorithm. {This is the case when the user is located in-between the respiratory cloud and the base station}. The three-bounce paths do not suffer from such a performance drop as they do not face the issue of colinearity of rays in this configuration.

\section{Challenges and Future Directions}

The development of {violent expiratory} aerosol cloud tracking based on the terahertz spectrum needs more development and analysis despite the promising concept and results shown in this paper. Below we list and discuss some of these challenges, and potential future directions of this technology for integration with the envisioned 6G infrastructure and new types of services for telecoms infrastructures.

\subsection{Moving Indoor Objects}

As discussed in this paper, {violent expiratory} aerosol clouds dissipate over the air as their droplets diffuse based on random motion towards other indoor areas, surfaces and floors. However, the dissipation of the {violent expiratory}  aerosol cloud depends also on the airflow present indoors, as jet streams and turbulent sources will affect how the aerosol cloud dissipates. Jet streams can move the  aerosol cloud faster to other areas depending on its source, potentially removing the {violent expiratory} aerosol cloud from the indoor space altogether. Turbulent sources from moving people and objects will also interfere with the cloud dissipation, either abruptly dislocating it and breaking it into multiple smaller ones. Since we consider the static dissipation of clouds, it is unclear how the scanning techniques can cope with the airflow effects on the {violent expiratory} aerosol cloud. This requires a deep characterization of the sources of jet streams and turbulences and the system response on those, as airflow properties such as velocity and movement patterns need to be coupled with existing models in the {violent expiratory} aerosol cloud dissipation. Even though there is existing literature on this topic, the need is for actual studies on the patterns of jet streams and turbulences indoors with the accuracy of the cloud tracking system and what adaptation is needed to cope with scaling airflow effects. As for the challenge of distinguishing between {violent expiratory}  aerosol and other sources of aerosol (e.g. a kettle), we envision an extension of the work that uses the characteristics of aerosol events (speed, plume shape, distribution of temperature and humidity)  {and its temporal dynamics that influence the detection of the respiratory cloud by our proposed technique, that is analyzed through machine learning algorithms}. 

\subsection{Environment Effects on Multiple Ray Scanning Techniques}

Since the propagation of terahertz beams depends on the air content in a given indoor scenario, the environmental factors that affect signal transmission will also impact the multiple ray scanning method used to track {violent expiratory} aerosol clouds. Besides the airflow effects already discussed in the previous section, humidity is a leading factor in terahertz signal degradation that will cause poor signal propagation at certain levels. The depth of the room's scanning technique depends on high-quality signals that will pass through {violent expiratory} aerosol clouds to be detected. High humidity levels mean that the signal will be degraded so much that the  aerosol cloud will disintegrate quickly, which will add new challenges in accurate detection. 



\subsection{Increased OPEX AND CAPEX Costs}

The efficiency of the sensing should determine the impact on the OPEX and CAPEX of the wireless network infrastructure. When scaling up the room's dimension, sensing will largely depend on the constant THz signal radiation in multiple locations so that the signal strength is enough to reach a meaningful recording of the {violent expiratory} aerosol cloud. Sensing efficiency also depends on the number of reflections points on walls. Compared to the planned infrastructure for 6G, there are no added CAPEX costs since the integration of terahertz base stations and IRS are an ongoing development for 6G. However, in terms of OPEX, energy consumption may increase based on the level of continuous sensing that is performed. 
It will be interesting to know how much energy is increased and what techniques can be developed to provide continuous monitoring without increasing the energy consumption.

\subsection{New Telecoms Services}

The most significant advantage of the proposed solution is the ability to use the wireless infrastructures for sensing purpose. This proposal will not rely on novel sensing hardware but will make use of the already available infrastructure. Sensing thus becomes an added commodity, whereby sensing not only {violent expiratory} aerosol clouds but other airborne agents become available. Gases and chemicals in high concentrations can also interact with the terahertz signals so that multiple agents can also be sensed \cite{chaccour2022seven}. Future telecom providers can offer new services with this sensing data. For example, mobile users that are located close to polluted or toxic areas can receive data relevant to specific gases that they are  exposed to and what levels. The inclusion of this service and the impact on telecom providers to offer sensed data that impacts on health can lead to new immeasurable benefits for society. We hope that our solution will make the IRS rollout faster with the added benefit of novel non-communication use.

\section{Conclusion}
The role future networks will play in {supporting} public health
will become increasingly pertinent in the coming decades. Experiences from the COVID-19 pandemic are emphasizing the need for a coordinated response and pervasive sensing of reliable information.  Our vision of using IRS and terahertz communications in 6G for detection and tracking of {violent expiratory} clouds recognises the unused potential of high-frequency communication systems for not only transmitting high speed data with high bandwidth, but also for a new form of sensing that will help public health.

An expansive and
ambitious vision of 6G networks is championed, whereby its success is measured by the human-centric
services it can provide to its users. The {violent expiratory} cloud detection service presented shows that
with the right design vision, 6G networks can be more than just a tool to exchange information,
and instead can become intelligent entities that actively participate in enriching and safeguarding
peoples’ lives.


%





\ifCLASSOPTIONcaptionsoff
  \newpage
\fi



\bibliographystyle{IEEEtran}

\vfill

\begin{IEEEbiographynophoto}{Harun \v{S}iljak}
(M'15, SM'20) graduated from the Automatic Control and Electronics Department, University of Sarajevo (B.E. 2010, M.E. 2012) and the International Burch University Sarajevo (Ph.D. 2015). He is an assistant professor in embedded systems at Trinity College Dublin, Ireland, working on complex systems and non-conventional computation, communications and networks.
\end{IEEEbiographynophoto}

\begin{IEEEbiographynophoto}{Michael Taynnan Barros}
(M'16) is a Lecturer at the School of Computer Science and Electronic Engineering, University of Essex, UK, and a Marie Skłodowska-Curie Individual Fellow at the BioMediTech Institute of the Tampere University, Finland. He received his PhD in Telecommunication Software at the Waterford Institute of Technology, Ireland, in 2016. 
\end{IEEEbiographynophoto}

\begin{IEEEbiographynophoto}{Nathan D'Arcy} obtained his master degree in Electronic Engineering from Trinity College Dublin in 2021. His professional interests include quantum engineering and machine learning.
\end{IEEEbiographynophoto}

\begin{IEEEbiographynophoto}{Daniel Perez Martins}
(STM'05, M'19) is a Postdoctoral Researcher and the Technical Lead of the Biomedical Nano and Molecular Telecommunications Team at Walton Institute. His research concentrates on the modeling and analysis of conventional and nanoscale communications systems. He received his PhD from the Waterford Institute of Technology, Ireland, in 2019. 
\end{IEEEbiographynophoto}

\vskip -1.2cm
\begin{IEEEbiographynophoto}{Nicola Marchetti}
(M'13, SM'15) is Associate Professor at Trinity College Dublin. He received the PhD in Wireless Communications from Aalborg University in 2007, the M.Sc. in Electronic Engineering in 2003, and the M.Sc. in Mathematics in 2010. His research interests include radio resource management, self-organising networks, and signal processing. 
\end{IEEEbiographynophoto}
\vskip -1.2cm
\begin{IEEEbiographynophoto}{Sasitharan Balasubramaniam}
 received his
Bachelor (electrical and electronic engineering) and Ph.D.degrees from the University of Queensland in 1998 and 2005, respectively, and his Master’s (computer and communication engineering) degree in 1999 from Queensland University of Technology. His current
research interests include the Internet of Nano Things and molecular communication.
\end{IEEEbiographynophoto}

\end{document}